\input{aipcheck}

\documentclass[
    ,final            
  ]
  {aipproc}

\layoutstyle{6x9}

\begin{document}
\rightline{SPhT--T06/031}

\title{QCD at the Dawn of the LHC Era}

\classification{11.15.-q,12.38.-t}
\keywords      {Perturbative QCD, twistors, shower Monte Carlo}

\author{David A. Kosower}{
  address={Service de Physique Th\'eorique\\CEA--Saclay\\
           F--91191 Gif-sur-Yvette cedex\\France}
}

\begin{abstract}
Precision computations of QCD and QCD+electroweak emission scattering
processes will play an important role in understanding experimental
data from the LHC, and thereby in uncovering new physics beyond the
Standard Model.  Recent years have seen substantial progress in
expanding our ability to perform and implement such calculations.
I discuss several different topics: the resolution of seeming discrepancies 
between heavy-quark predictions and measurements; the ongoing program
of next-to-next-to-leading order calculations; and also progress in
merging the parton-shower and matrix-element calculations.  I will
also outline some of the techniques emerging from twistor-space string
theory and its intellectual progeny.
\end{abstract}

\maketitle


\section{Introduction}

As the particle-physics community looks forward
to the start of the experimental program at the LHC, 
we should remember that essentially every aspect of particle physics 
at a hadron collider involves perturbative QCD.  Calculations in QCD are therefore
an essential part of any theoretical support to this program.  These include both
predictions of signals and backgrounds, as well as precision calculations needed for 
using $W$s for luminosity measurements and for extraction of the parton distribution
functions.

The requirements for QCD calculations present the theorists amongst us
with a serious challenge, because the coupling is not small and its running 
is important.  This implies that we have to deal with lots of jets, and to higher order if we want precision.
  Furthermore, because processes can 
involve 
very different scales, logarithms can be substantial, and we have to resum them.

The technology for required perturbative calculations has advanced greatly over the 
last two decades, but the calculations are still difficult.  Accordingly, we must settle
for a partial attack on aspects of many processes.  Indeed, for some processes, such as
the $W+4$~jets processes giving the dominant background to top quark detection in the
lepton+jets channel, only the most basic perturbative QCD prediction has been available to date.

There are two basic approaches that have been pursued by theorists to computing
collider processes: (a) general parton-level calculations (b) parton-shower calculations, colloquially known as Monte Carlos.
In the parton-level calculations, one computes all contributions to a process of interest up to some fixed order in perturbation
theory.  With the required matrix elements in hand, one can write a numerical jet program, computing general observables.  One can vary
the jet algorithm, and implement arbitrary cuts. 
Theorists like this approach, because one computes systematically to higher order and higher multiplicity in perturbation theory.
So long as one avoids observables with large ratios of scales (which can lead to large logarithms needing resummation), the results
are then accurate up to higher-order corrections in $\alpha_s$.  Experimenters dislike this approach, because given the limited number of partons in the final state, one cannot study a process event-by-event; one can only compare distributions of events.  The limited
number of partons also makes it harder to implement faithfully the intricate cone jet algorithm typically used by hadron-collider
experiments in their analyses.  

In parton-shower calculations such as implemented in P{\sc ythia\/}~\cite{Pythia} or {\sc Herwig\/}~\cite{Herwig}, 
one has in the past started with a basic $2\rightarrow1$ or
$2\rightarrow2$ perturbative process, and added additional parton radiation using the collinear approximation.  This 
produces a detailed partonic picture on an event-by-event basis.  It can also be used to compute general observables, and in
addition can have (and is usually used with) a model of hadronization folded on, so that a detailed simulation of the detector is
possible.  For this reason, experimenters like this approach.  Theorists didn't like the traditional parton-shower approach as much,
because it provides only a leading-log estimate (perhaps with some next-to-leading log contributions) of observables.  
Accordingly, this approach was not systematic, and not amenable to systematic improvement.  The terms not
included are hard to estimate numerically, and vary widely across regions of phase space.  A particular example is the uncontrolled
underestimation of wide-angle radiation.

One of the exciting developments in recent years, to which I return below, is the start of a convergence between these different
approaches, through incorporation of higher-multiplicity and higher-order corrections into the parton-shower approach.

The incorporation of higher-order corrections will require the computation of a large number of one-loop amplitudes.
At tree level, there are a number of general-purpose programs which can mechanically and automatically produce
matrix elements for a wide variety of processes required for collider physics.  We are still far from such a situation for
one-loop corrections, because the computational complexity of traditional diagrammatic methods is just too great.  The last
year has seen dramatic advances on this front, thanks to the addition of ideas based on a {\it twistor-space string duality\/} to
an older non-conventional technique, the unitarity-based method for loop computations.

\section{Heavy-Quark Predictions}

Before turning to recent advances in these topics, let's first review the resolution of an old puzzle in perturbative
QCD, that of heavy-quark production at hadron colliders.   A long series of comparisons between experimental measurements
at CDF~\cite{CDF} and D0~\cite{D0} and theoretical predictions~\cite{QuarkTheory} for 
$b$ quark production appeared to show a factor-of-two discrepancy, with the theory
underestimating production.  This issue has now been laid to rest by Cacciari, Frixione, Mangano, Nason, and Ridolfi~\cite{CFMNR}. There are
several ingredients in the resolution.  The first is a proper theoretical treatment of $b$ quark fragmentation into bottomful hadrons,
as well as use of fragmentation functions extracted from $e^+ e^-$ data.  They also incorporated log resummation (with proper matching).
A comparison to directly-predictable data --- the production of $J/\psi$s from $B$ decay --- rather than `inclusive $b$' production also
plays a role.
Small changes in parton distribution functions and $\alpha_s$ complete the picture.

\section{Advances in the Parton-Shower Approach}

In the parton-shower approach as conceived in the late '70s--mid '80s, jet radiation is approximated using the collinear
approximation.  One starts with a $2\rightarrow2$ process, gives each outgoing parton a mass of order the $Q^2$ of the subprocess,
and lets it decay using the appropriate Altarelli--Parisi splitting kernel.  Energy-momentum conservation is adjusted at the end.
This approximation is suitable for narrow jets, where it gives the leading-log contributions
(enhancements include next-to-leading log contributions
as well).  In practice, it appears to work well for jets as measured by real-world experiments.  The same approximation is
also used in the basic parton-shower approach for wide-angle radiation, in particular for additional jets.  Here, the approximation
is uncontrolled; and not too surprisingly, shapes of distributions can show significant differences from parton-level calculations.
The basic parton-shower approach also does not allow for a systematic improvement to take into account higher orders in perturbation
theory, as is possible and increasingly practiced in parton-level calculations.  

Are there ways to solve these two problems, improving
parton shower programs both to higher multiplicity as well as higher order?
These questions have attracted a great deal of theoretical attention in recent years, and while the problem cannot yet be considered
solved, much progress has been made.  One might imagine attacking the higher-multiplicity problem by showering off higher-multiplicity
parton configurations instead of or in addition to the basic $2\rightarrow 2$ ones.  This raises a double-counting problem, however,
because both the higher-multiplicity parton matrix elements and the showering off lower-multiplicity ones will contribute in
the near-collinear region, that is to intra-jet radiation.  One has to separate these two contributions.  In addition, one must confront
the problem that the near-collinear contributions of higher-multiplicity parton matrix elements are infrared divergent.  
Catani, Krauss, Kuhn, and Webber~\cite{CKKW} proposed a solution to these problems.  The basic approach is as follows: start by generating a
different fixed-order configurations, 
using a $k_T$ separation to eliminate the infrared divergences in the higher-multiplicity partonic matrix
elements.  Each partonic configuration is assigned a branching history, as though it came from showering, and reweighted with appropriate
Sudakov factors.  Finally, showering is started at the $k_T$ at which the fixed-order contribution was cut off.
The CKKW proposal has now been implemented~\cite{CKKWImplementation} in the {\sc Sherpa\/}~\cite{Sherpa} generator.

If we wanted to improve a parton-shower calculation by adding loop corrections, or equivalently improve our ability to simulate
a next-to-leading order calculation at the fine-grained level of a detector simulation, we would face a similar kind of problem.
Simply adding parton showers to an NLO calculation would  double-count virtual contributions.

Frixione and Webber proposed a solution to this problem, implemented in the MC@NLO program~\cite{MC@NLO}.  They subtracted the double-counted
terms generated by the first branching in the shower.  It has since been implemented for a number of other processes, including
heavy-quark production.  The approach is
not completely general, and does require specific calculations of terms for each new process.   Work continues on more general
approaches, such as one based on dipole subtraction~\cite{DipoleShower}.  This area seems likely to produce further advances in coming years.
 
\section{Higher-Order Calculations}

Leading-order (LO) parton-level calculations require only tree-level matrix elements.  There are several programs which
serve as black boxes (CompHEP~\cite{CompHEP}, HELAC/PHEGAS~\cite{Helac}, 
GRACE~\cite{Grace}, Amegic~\cite{Amegic},
MadEvent~\cite{MadEvent}, ALPGEN~\cite{Alpgen}), and allow the calculation of a wide variety of such matrix elements both in the Standard Model
and popular extensions of it.  Such leading-order calculations give a basic prediction of event shapes, but because
of the sensitivity to the renormalization scale in the coupling, do not give a reliable quantitative prediction. 
They also model jets via lone partons.  This completely
misses sensitivity of experimental measurements to the jet definition, for example to such parameters as the jet cone size or
minimum transverse energy.
The current state of the art in parton-level calculations for most processes 
is a next-to-leading order (NLO) calculation.  Such calculations incorporate one-loop corrections.  These introduce
an explicit dependence on the renormalization scale, which compensates that present in the strong couple $\alpha_s$, and thereby
leads to a more stable, and quantitatively reliable, prediction.  In addition to the one-loop corrections, radiative emission
contributions must also be included.   These real-emission contributions introduce a dependence on the jet-definition parameters, thereby
giving a first theoretical approximation to this sensitivity.  NLO calculations do have a greater technical complexity than those at LO.
The one-loop corrections have infrared divergences, which are conventionally regulated via
a dimensional regulator.  The phase-space integrals over the real-emission contributions are also infrared divergent.  We want to
perform the phase-space integrals numerically while isolating the divergences, and computing the divergences analytically.  General
schemes to do this in a process-independent way have been around for a decade~\cite{Slicing,FKS,DipoleSubtraction}.
For a new process, the only new calculation one has to do is of the corresponding
one-loop amplitude.  (The
alternative of computing the divergent parts numerically as well has also been tried out~\cite{PurelyNumerical}.)

For true precision predictions, at the percent level, NLO does not suffice; one must go
an order beyond, to next-to-next-to-leading order (NNLO).
  These will certainly be required for the $W$+jet process, especially if one is to use it as
for measuring luminosity.  Anastasiou, Dixon, Melnikov, and Petriello~\cite{ADMP}
 have done semi-analytic calculations of the $W$ rapidity distribution.  
While a fully numerical program will be required to implement experimental cuts, their calculation does show that perturbation theory
is settling down nicely at this order, and that obtaining the required accuracy is feasible.  NNLO calculations should also allow
better prediction of the dependence on jet definitions, as well as a first quantitative estimate of the remaining theoretical 
uncertainties.  Assembling an NNLO prediction for, say, $W$+~jet production is a technically challenging task.  It is underway,
but will have taken a small community many years of hard work to accomplish by the time it is complete.  One needs two-loop matrix
elements, in this case for $2 {\rm\ partons}\rightarrow W+{\rm parton}$~\cite{TwoLoopME}.  
Such computations were made possible several years ago by
dramatic advances in calculating multi-scale two-loop integrals~\cite{TwoLoopIntegrals}.  In addition, one needs the 
one-loop $2 {\rm\ parton}\rightarrow W+2{\rm\ parton}$
matrix elements, which were computed a decade ago~\cite{Zqqgg,OneLoopME}; and $2 {\rm\ parton}\rightarrow W+3{\rm\ parton}$ tree-level 
matrix elements, which have been
known since the 1980s~\cite{TreeLevelME}.  
One needs to extract the singular parts of the two-loop amplitudes, as well as the singular terms in the one-loop
matrix elements and the doubly-singular terms the tree-level matrix elements.  The latter two types have to be integrated over phase
space in order to determine the double-real emission and the mixed real-virtual singularities.  The cancellation of singular terms
has to be demonstrated, and the remaining finite terms organized into a computer program.  Hadron-collider processes are harder,
because of the need to handle initial-state factorization.  It is somewhat simpler --- but still technically challenging ---
 to start with the $e^+ e^-\rightarrow 3{\rm\ jets}$
process, and that is what the G$^3$ collaboration~\cite{GGG} is doing.

Hadron collider processes require the NNLO corrections to the Altarelli--Parisi splitting kernel.  Fortunately,
the last couple of years have seen the completion of a landmark calculation, several years of effort by Moch, Vermaseren, and Vogt, of this quantity~\cite{MVV}.  The calculation confirms the stability of perturbation theory, and will be essential to any precision physics to be done at LHC.  It's already being incorporated into the evolution of the pdfs as used for example in their extraction.  It's also worth mentioning that the calculation is even of interest to string theorists, since one can use it to extract anomalous dimensions relevant to the so-called AdS/CFT correspondence~\cite{KLOV}.

\section{Lessons from Twistor Space}

The last two years have seen dramatic new developments in our understanding of
gauge theories.  These developments have also started to contribute practical
advances relevant to perturbative QCD computations for the LHC.  The roots of these
developments lie in Penrose's invention of twistor space~\cite{Penrose} over thirty years ago, and
also in Nair's representation~\cite{Nair} of the simplest gauge-theory amplitudes, maximally
helicity-violating (MHV) ones, as correlators on `baby twistor space' or CP${}^1$.
The developments flow from Witten's conjecture~\cite{Witten} that ${\cal N}=4$ supersymmetric
gauge theory is dual to a topological string theory with a CP${}^{3|4}$ (that is, superprojective twistor) 
target space.

This duality is of a novel kind.  Unlike the strong--weak coupling duality
relating the same supersymmetric gauge theory to type IIB string theory on
an AdS${}_5\times S^5$, this conjectured duality is between the gauge theory in
the perturbative, weak-coupling regime and a string theory in the {\it same\/}
regime.  The conjecture has led to new representations of tree-level amplitudes,
some of which also have a field-theory interpretation, and some of which do not appear to.
It has revealed that tree (and plausibly, loop) amplitudes satisfy unexpected differential
equations.  The general lesson seems to be that there is additional structure or
symmetry in gauge theories beyond what is manifest in the Lagrangian.

\def\tlambda{{\widetilde\lambda}}
We can get to twistor space from the spinor-helicity
representation~\cite{SpinorHelicity} of QCD amplitudes that was developed in the 1980s and has been
an invaluable tool in perturbative computations ever since.  If we consider a
massless four-momentum $k$, we can form a $2\times2$ complex matrix by contracting
it with the Pauli sigma matrices, $K_{a\dot a} = \sigma_{a\dot a}^\mu k_\mu\,$.
The massless condition implies that $\det K = 0$, so that we can write $K$ as the
tensor product $\lambda_{a\vphantom{\dot a}} \tlambda_{\dot a}$.  The complex
two-vectors $\lambda$ and $\tlambda$ are spinors under $SU(2)$.  If we now
Fourier-transform $\tlambda$, calling the conjugate variable $\mu$, while leaving $\lambda$ alone,
we obtain a complex four-vector $Z^I = (\lambda,\mu)$.  Similarly, polarization vectors can
be written in terms of either $\lambda$ or $\tlambda$ and a conjugate spinor related to a lightcone `reference' vector.
Defining an equivalence relation whereby
$Z^I \equiv Y^I \tau$, related to the invariance (up to a phase) of helicity amplitudes under rescalings
of the spinors, defines a complex projective space, CP${}^3$.  The supertwistor space
in Witten's construction is the supersymmetric generalization obtained by adding four
fermionic coordinates.

Nair's representation appears in Witten twistor-string theory as the statement that
the MHV amplitudes are supported on CP${}^1$s (called `lines') embedded in CP${}^{3|4}$.
An $n$-point MHV amplitude has $n-2$ gluons of positive helicity, and $2$ gluons of negative helicity.
(Amplitudes with fewer negative helicities vanish.)  Amplitudes with more negative-helicity
gluons are more complicated, and are represented by higher-degree curves (that is, embeddings
of the basic CP${}^1$).  Roiban, Spradlin, and Volovich~\cite{RSV} have shown that
this representation does indeed yield the same answers as previous field-theory based computations.

What is more surprising is that, in a manner analogous to that of a contour integral, the twistor-string
integral giving rise to these amplitudes can also be evaluated on {\it degenerate\/} higher-degree curves,
in particular curves made up of intersecting line segments.  This representation is not only simpler than
expected in twistor space, but also has a very simple field-theory interpretation, in the form
of the Cachazo--Svr\v{c}ek--Witten (CSW) construction~\cite{CSW}.  In it, one builds up amplitudes from vertices
which are simple off-shell continuations of the MHV amplitudes, connected by scalar propagators.  The vertices
have exactly two negative-helicity gluons, and an arbitrary number of positive-helicity ones.  The helicity
projectors ordinarily part of the propagator here are effectively merged into the vertices.  For any given
process, one includes all diagrams built out of these vertices.  Each diagram gives rise to a single factor,
and so one can immediately write down an analytic expression in terms of spinor products for a given amplitude.

Witten's conjecture motivated a series of one-loop computations of amplitudes~\cite{NMHVComputations,NeqFourNMHV}
 to check their twistor-space
structure.  These amplitudes are indeed likely to have a simple structure in twistor space~\cite{OneLoopStructure,BCFcoplanar,NeqFourNMHV},
 though the twistor string itself is not completely understood at one loop~\cite{BerkovitsWitten}.  More suprisingly, however, these one-loop
computations led to another novel representation of {\it tree\/} amplitudes.  This came about through the
infrared-divergent parts of the one-loop amplitudes, which are proportional to the corresponding tree amplitudes.
These turned out to have a {\it simpler\/} representation~\cite{RSVNewTree} than would be obtained by direct calculation (or
indeed by calculation from the CSW representation).  The price of the compactness was the introduction of
spurious singularities, present term-by-term but cancelling in the amplitude as a whole.  (Such spurious
singularities arise naturally in loop computations.)  This representation can also arises, as it turns out,
from a novel kind of recursion relation, an {\it on-shell\/} one~\cite{BCFRecursion}.  That is, an amplitude is written as a sum
of terms, each term consisting of an ordinary propagator multiplied by two {\it on-shell\/} amplitudes.  This
is quite unlike a conventional field theory representation, where the internal lines would necessarily be
off shell.  Such a representation, maintaining the on-shell conditions for all legs along with momentum
conservation, is made possible by the use of {\it complex\/} rather than real momenta.

Even more surprisingly, Britto, Cachazo, Feng and Witten (BCFW) showed that
these on-shell recursion relations have a simple and very general
proof, relying only on factorization and complex analysis.  Accordingly, the same
approach can be applied to amplitudes in many field theories, including gravity~\cite{Gravity}; massive theories~\cite{Massive};
and even more generally, to the rational functions which are the coefficients of some of the integrals that
appear in one-loop amplitudes~\cite{IntegralCoefficients}.  Moreover, a simple variant also
yields the CSW representation as an on-shell recursion relation~\cite{Risager}.

These on-shell recursion relations are an example of how knowledge of the physical
properties of amplitudes (in this case, factorization) can be used as a tool to compute
them.  This is similar in spirit to an older non-conventional method, the unitarity-based
method for loop calculations~\cite{UnitarityMethod}.  While factorization suffices for tree amplitudes, which are
rational functions of spinor products, at loop level one needs additional information.  It turns out that
in supersymmetric theories, unitarity (that is, knowledge of the four-dimensional on-shell tree amplitudes
which make up the cuts) is sufficient to compute the complete amplitude.  Because of the worse
ultraviolet power-counting properties of nonsupersymmetric theories, however, there are rational
terms in their amplitudes that are not determined by four-dimensional unitarity.  

While $D$-dimensional unitarity can be used to determine these terms ($D=4-2\epsilon$)~\cite{SelfDual}, the required
amplitudes are more complicated.  An alternative is to use factorization to determine these terms, as was
done in a computation of the matrix elements for $V\rightarrow q{\overline q}gg$~\cite{Zqqgg}.  There, use of
factorization required guessing an ansatz, and did not provide a systematic approach to obtaining the rational
terms.  An on-shell recursion relation, similar to BCFW's, does provide such a systematic method~\cite{OneLoopRecursion}.  
In combination
with the unitarity-based method, it provides a powerful and efficient method for computing the sorts of
one-loop amplitudes that are required for the LHC.  In the combined `unitarity-bootstrap' approach, the cut-containing terms (the logarithms,
polylogarithms, and $\pi^2$) are determined using four-dimensional unitarity.  A recursion relation determines the rational
terms, and removing terms which are double-counted yields the complete QCD amplitude.
This approach has already been applied to six-point amplitudes which have
proven too difficult for conventional brute-force diagrammatic methods, and to an infinite series of $n$-point amplitudes~\cite{AllnMHV},
a calculation which is simply not possible with conventional techniques.






\begin{theacknowledgments}
It is a pleasure to thank Carola Berger, Zvi Bern, Lance Dixon, and Darren Forde for their collaboration on the
unitarity and unitarity-bootstrap methods discussed above.  
\end{theacknowledgments}



\bibliographystyle{aipproc}   




\end{document}